\documentclass[letterpaper,journal]{IEEEtran}
\usepackage{amsmath,amsfonts,amssymb}
\usepackage{algorithmic}
\usepackage{array}
\usepackage[caption=false,font=normalsize,labelfont=sf,textfont=sf]{subfig}
\usepackage{textcomp}
\usepackage{stfloats}
\usepackage{url}
\usepackage{verbatim}
\usepackage{graphicx}
\def\BibTeX{{\rm B\kern-.05em{\sc i\kern-.025em b}\kern-.08em
    T\kern-.1667em\lower.7ex\hbox{E}\kern-.125emX}}
\usepackage{balance}
\usepackage{bm}
\usepackage{cite}
\usepackage{mathtools}
\usepackage{subfig}
\usepackage{lettrine}
\usepackage{color}
\usepackage[utf8]{inputenc}
\usepackage{verbatim}

\begin{document}

\title{DUGC-VRNet: Joint VR Recognition and Channel Estimation for Spatially Non-Stationary XL-MIMO}
\author{Jinhao~Nie,~\IEEEmembership{Graduate~Student~Member,~IEEE}, %
        Guangchi~Zhang,
        Miao~Cui,
        Hao~Fu,
        and Xiaoli~Chu,~\IEEEmembership{Senior~Member,~IEEE}%
        
        \thanks{
        This work was supported in part by Guangdong Basic and Applied Basic Research Foundation under Grant 2026A1515011208 and in part by Guangdong Overseas Distinguished Teachers Project under Grant MS202500028.
        
        Jinhao Nie, Guangchi Zhang, Miao Cui, Hao Fu are with the School of Information Engineering, Guangdong University of Technology, Guangzhou 510006, China (e-mail: 2112403060@mail2.gdut.edu.cn; \{gczhang, cuimiao, ecejasonhaofu\}@gdut.edu.cn). Xiaoli Chu is with the School of Electrical and Electronic Engineering, The University of Sheffield, S1 3JD Sheffield, U.K. (e-mail: x.chu@sheffield.ac.uk). Corresponding authors: G. Zhang and M. Cui.
        }
}


\maketitle
\begin{abstract}
In this letter, we address spatially non-stationary near-field channel estimation for extremely large-scale multiple-input multiple-output (XL-MIMO) systems with a hybrid combining architecture. One key challenge in the considered problem lies in that conventional channel estimation algorithms typically struggle to effectively identify and adapt to the partial antenna visibility caused by varying visibility regions (VRs), thereby compromising estimation accuracy. To perform joint VR recognition and channel estimation, we integrate a deep unfolding network (DUN) with a graph convolution network (GCN), leading to a Deep Unfolding and Graph Convolution coupled, Visibility Region Aware Network (DUGC-VRNet). By leveraging the channel’s graph structure, the GCN infers and feeds back VR information to dynamically guide the DUN's updates, thereby enhancing reliable channel estimation under spatial non-stationarity. To reduce DUGC-VRNet's complexity, we apply weight pruning to obtain a lightweight network. Simulation results demonstrate that the DUGC-VRNet and its pruned variant achieve superior channel estimation and more accurate VR recognition under spatially non-stationary conditions.
\end{abstract}

\begin{IEEEkeywords}
    XL-MIMO channel estimation, spatial non-stationary, visibility region recognition, graph convolution network, weight pruning.
\end{IEEEkeywords}

\section{Introduction}
\IEEEPARstart{E}{xtremely} large-scale multiple-input multiple-output (XL-MIMO) has emerged as an enabling technology for 6G to achieve substantial gains in spectral efficiency and spatial resolution by employing massive antenna arrays. It is typically deployed in high-frequency bands, such as millimeter-wave (mmWave) bands \cite{ref1}. The large array apertures and high carrier frequencies often place XL-MIMO communications in the near field, where spherical-wave propagation must be considered, and the channel characteristics depend on both angle and distance \cite{ref2}. Concurrently, the large antenna aperture induces spatial non-stationarity, meaning that different antennas experience non-uniform fading. This phenomenon is captured by the visibility region (VR), which indicates the subset of antennas effectively seen by each user \cite{ref5}. 

XL-MIMO channel estimation under spatial non-stationarity remains a challenge, which has attracted several investigations. Chen et al. reformulated the spatially non-stationary channel as a set of spatially stationary sub-channels and estimated them using a group time block code (GTBC), including GTBC-based simultaneous orthogonal matching pursuit (GP-SOMP) and GTBC-based polar-domain simultaneous iterative gridless weighting (GP-SIGW) \cite{ref4}. Zhang et al. \cite{ref3} proposed a two-level OMP (TL-OMP) algorithm and a list-type channel estimation (LT-CEM) algorithm for joint channel estimation and VR recognition. However, the above methods rely on hand-crafted sparse dictionaries to model spatial non-stationarity, which increases pilot overhead and reduces robustness.

To address these issues, Wu et al. \cite{ref15} exploited the structural non-stationarity of the channel and performed estimation via a fixed-rank manifold gradient descent (FRM-GD) algorithm. In parallel, several joint VR detection and channel estimation methods have been developed under Bayesian inference frameworks. Tang et al. proposed the VR detection-oriented message passing (VRDO-MP) \cite{ref14} and the three-layer generalized approximate message passing (TL-GAMP) \cite{ref13} schemes, while Xu et al. \cite{ref16} introduced an alternating maximum a posteriori (MAP) strategy for uniform planar array (UPA) based systems. Though these approaches improve pilot efficiency and robustness, they typically rely on predefined statistical assumptions or hand-crafted priors, which may limit their adaptability in complex, spatially non-stationary environments.

Recently, deep learning (DL)-based methods have been proposed for XL-MIMO channel estimation. Ye et al. \cite{ref11} utilized graph neural networks to exploit the channel's graph structure for massive MIMO channel estimation. Gao et al. \cite{ref8} employed a convolutional neural network (CNN) to estimate XL-MIMO channels. Zheng et al. \cite{ref10} proposed a model-driven iterative super-resolution channel estimation network (MDISR-Net) and demonstrated that deep unfolding networks (DUNs) can effectively estimate near-field XL-MIMO channels in hybrid radio frequency (RF) array architectures. However, the above methods do not explicitly model the spatial non-stationarity of XL-MIMO, leading to biased estimates in complex spatially non-stationary environments.

In this letter, we integrate a DUN with a graph convolution network (GCN) for VR feedback, thus obviating the need for hand-crafted sparse dictionaries and propose a novel Deep Unfolding and Graph Convolution coupled Visibility Region Aware Network (DUGC-VRNet). Different from existing dictionary-based methods, Bayesian inference approaches, and deep unfolding schemes, DUGC-VRNet enables joint VR recognition and channel estimation, where the GCN exploits the channel’s intrinsic graph structure to infer a VR mask for VR recognition, and the proposed feedback mechanism feeds the GCN-inferred VR information back to the DUN to improve channel estimation under spatial non-stationarity. In addition, we apply global weight pruning to compress DUGC-VRNet into a lightweight model with minimal performance loss. Simulation results verify that DUGC-VRNet consistently outperforms existing algorithms in both channel estimation and VR recognition with limited pilot overhead and at low SNR.

\section{System Model and Existing Channel Estimation Method}
\subsection{System Model}
We consider the uplink of a narrowband XL-MIMO system operating at mmWave frequencies. A base station (BS), equipped with an $N$-antenna uniform linear array (ULA), serves $K$ single-antenna users. The antenna spacing is $d = \lambda/2 = c/(2f_c)$, where $f_c$ is the carrier frequency and $\lambda$ is the wavelength.

Large apertures and high carrier frequencies place XL-MIMO in the electromagnetic near field \cite{ref2}, where spherical-wave propagation and spatial non-stationarity must be considered. Specifically, only a subset of the BS array, defined by a visibility region (VR), is visible to each user. The channel vector between user $k$ and the BS is expressed as \cite{ref3}
\begin{equation}
\label{deqn_ex1a}
\mathbf{h}_{k} = \gamma_{k} \, \mathbf{b}(\theta_k, r_k) \odot \mathbf{u}_k,
\end{equation}
where $\gamma_{k} = \sqrt{N} \frac{\lambda}{4\pi r_k} e^{-j \frac{2\pi}{\lambda} r_k}$, and $r_k$ and $\theta_k$ are the distance and angle of arrival from user $k$ , respectively. The near-field steering vector $\mathbf{b}(\theta_k, r_k)$ is given by
\begin{equation}
\label{deqn_ex2a}
\mathbf{b}(\theta_k, r_k) = \frac{1}{\sqrt{N}} \left[ e^{-j \kappa (r_k^{(1)} - r_k)}, \ldots, e^{-j \kappa (r_k^{(N)} - r_k)} \right]^{\mathrm{T}},
\end{equation}
where $\kappa = 2\pi / \lambda$, and $r_k^{(n)} = \sqrt{r_k^2 + \delta_n^2 d^2 - 2 r_k \delta_n d \theta_k}$ is the distance from user $k$ to the $n$-th antenna with $\delta_n = (2n - N - 1)/2$. The VR mask $\mathbf{u}_k \in \{0,1\}^N$ represents the antenna visibility. Following \cite{ref3}, the ULA is divided into $S$ stationary subarrays, each with $N_s = N/S$ antennas. Let $\mathbf{u}_{k}' \in \{0,1\}^{S}$ be the subarray-level VR mask; then the antenna-level mask is $\mathbf{u}_k = \mathbf{u}_k' \otimes \mathbf{1}_{N_s}$, where $\mathbf{1}_{N_s}$ is an all-ones vector of length $N_s$.

Assuming orthogonal pilots, we focus on an arbitrary user and omit the index $k$. To estimate the channel, the user transmits $P$ pilot symbols over $P$ time slots. The BS employs a hybrid architecture with $N_{\mathrm{RF}}$ RF chains. Let $x_p = 1$ be the pilot symbol at slot $p$, and the received signal is
\begin{equation}
\label{deqn_ex3a}
\mathbf{y}_{p}=\mathbf{A}_p\,\mathbf{h} + \mathbf{A}_p\,\mathbf{n}_{p}\in \mathbb{C}^{N_{\mathrm{RF}} \times 1}, \quad p=1, \ldots, P,
\end{equation}
where $\mathbf{h} \in \mathbb{C}^{N \times 1}$, $\mathbf{A}_p \in \mathbb{C}^{N_{\mathrm{RF}} \times N}$ is the analog combiner at slot $p$ with constant-modulus entries $|\mathbf{A}_p(i,j)| = \frac{1}{\sqrt{N}}$, and $\mathbf{n}_{p} \sim \mathcal{CN}(\bm{0}, \sigma^2 \mathbf{I}_N)\in \mathbb{C}^{N \times 1}$ is the noise. By collecting the observations from the $P$ pilot slots, we obtain
\begin{equation}
\mathbf{\tilde{y}}=\mathbf A \mathbf h + \mathbf{\tilde{n}}=
\big[
(\mathbf A_1 \mathbf h)^{T},
\ldots,
(\mathbf A_P \mathbf h)^{T}
\big]^{T}
+ \mathbf{\tilde{n}} \in \mathbb{C}^{N_{RF}P \times 1} .
\label{eq4}
\end{equation}
where $\mathbf{\tilde{y}} =[\mathbf{y}_1^{T}\ldots,\mathbf{y}_P^{T}]^{T}$, $\mathbf{A} = [\mathbf{A}_1^{T},\ldots,\mathbf{A}_P^{T}]^{T}$, $\mathbf{\tilde{n}}
=\big[(\mathbf A_1 \mathbf n_1)^{T},\ldots,(\mathbf A_P \mathbf n_P)^{T}\big]^{T}$.

\subsection{Existing MDISR-Net-Based Channel Estimation Method}
The optimal estimation of $\mathbf{h}$ from $\mathbf{\tilde{y}}$ is given by  \cite{ref9},
\begin{equation}
\label{deqn_ex5a}
\begin{aligned}
\widehat{\mathbf{h}}=\underset{\mathbf{h}}{\arg\min} \frac{1}{2}\|\mathbf{\tilde{y}}-\mathbf{A} \mathbf{h}\|_2^2+\lambda \varphi(\mathbf{h}),
\end{aligned}
\end{equation} where $\varphi(\mathbf{h})$ denotes a regularization term capturing the characteristics of the near-field channel, and $\lambda$ is a positive parameter controlling this regularization. The hybrid RF architecture reduces the dimension of $\mathbf{\tilde{y}}$ from $N$ to $N_{\mathrm{RF}}$, rendering channel recovery an underdetermined problem that necessitates exploiting channel sparsity for accurate estimation. While conventional compressed-sensing approaches rely on hand-crafted dictionaries, these often need to be enlarged or supported by more pilots to accurately represent near-field and spatially non-stationary channels, thereby increasing complexity and overhead while offering limited estimation performance \cite{ref3,ref4}. MDISR-Net \cite{ref10} efficiently addresses this challenge. By introducing a prior channel $\mathbf{z}$ to decouple the data fidelity term $\frac{1}{2}\|\mathbf{\tilde{y}}-\mathbf{A} \mathbf{h}\|_2^2$ and the regularization term $\lambda \varphi(\mathbf{h})$, it learns a data-driven prior $\mathbf{z}$ for the near-field channel directly from training data. This approach resolves the underdetermined nature without relying on hand-crafted dictionaries. Specifically, in the $t$-th layer of MDISR-Net, it solves the following subproblems
\begin{align}
\mathbf{h}^{(t+1)} = & \underset{\mathbf{h}}{\arg \min } \frac{1}{2}\left\|\mathbf{\tilde{y}}-\mathbf{A} \mathbf{h}\right\|_2^2+\frac{\mu}{2}\left\|\mathbf{h}-\mathbf{z}^{(t+1)}\right\|_2^2,  \label{deqn_ex7a} \\
\mathbf{z}^{(t+1)}  =  & \underset{\mathbf{z}}{\arg \min } \frac{\mu}{2}\left\|\mathbf{h}^{(t)}-\mathbf{z}\right\|_2^2+\lambda \varphi(\mathbf{z}),  \label{deqn_ex8a}
\end{align}
where $\mu$ is a penalty parameter.

\section{DUGC-VRNET: Joint VR Recognition and Channel Estimation} 
We propose DUGC-VRNet, a $T$-layer unfolded network that integrates a DUN with a GCN. The key novelty is to recognize the VR and feed the derived VR back into the channel update, yielding VR-aware estimation. In the following, we present DUGC-VRNet’s architecture and principle, then describe its training, loss function design, and weight pruning.

\subsection{Principle of DUGC-VRNet}
DUGC-VRNet establishes a closed loop of VR recognition, feedback, and channel update, yielding interpretable VR outputs and reducing noise-induced bias for higher accuracy. Specifically, taking the $t$-th layer as an example, we have 
\begin{align}
\label{deqn_ex9a}
\mathbf{h}^{(t+1)}
&= \underset{\mathbf{h}}{\arg\min}\
   \frac12\!\big\|\mathbf{\tilde{y}}-\mathbf{A}\mathbf{h}\big\|_{2}^{2}\notag\\
&\quad+\frac{\mu}{2}\!\big\|\mathbf{W}\!\big(\mathbf{u}^{(t)}\big)^{\frac12}
   \!\big(\mathbf{h}-\mathbf{z}^{(t+1)}\big)\big\|_{2}^{2},
\end{align}
which can be interpreted as a weighted minimization problem, where the optimizer selects the $\mathbf{h}$ that yields the smallest weighted cost among all candidates. To encode the spatial non-stationarity, we employ a diagonal weight matrix $\mathbf{W}(\mathbf{u}^{(t)})\in\mathbb{R}^{N\times N}$ to inject the $t$-th VR mask $\mathbf{u}^{(t)}$ into $\left\|\mathbf{h}-\mathbf{z}^{(t+1)}\right\|_2^2$. The square-root form $\mathbf{W}(\mathbf{u}^{(t)})^{1/2}$ directly weights the discrepancy, and its diagonal elements $w_{ii}$ are given by 
\begin{equation}
\label{eq:w_ii_definition}
w_{ii}\!\big(\mathbf{u}^{(t)}\big)
= \sqrt{\,1+\beta\!\left(1-u_i^{(t)}\right)}\,,\; i=1,\ldots,N,
\end{equation}
where $\beta \gg 1$ is a penalty. For out-of-VR antennas, where the VR mask of the $i$-th antenna $u_i^{(t)} \approx 0$, $w_{ii} \gg 1$. Thus, any $\mathbf{h}$ with energy on these antennas faces much higher penalties in $\big\| \mathbf{W}(\mathbf{u}^{(t)})^{1/2}\!\big(\mathbf{h}-\mathbf{z}^{(t+1)}\big) \big\|_{2}^{2}$, thereby discouraging solutions that attribute such energy to VR energy. Conversely, for in-VR antennas ($u_i^{(t)} \approx 1$), $w_{ii} \approx 1$, resulting in weak penalties.

Due to the impact of spatial non-stationarity on $\mathbf{z}$, we impose a spatially weighted $\ell_1$ penalty $\big\|\big(\mathbf{1}-\mathbf{u}^{(t)}\big)\odot \mathbf{z}\big\|_{1}$:
\begin{align}
\label{deqn_ex11a}
\mathbf{z}^{(t+1)}
&= \underset{\mathbf{z}}{\arg\min}\;
   \frac{\mu}{2}\,\big\|\mathbf{h}^{(t)}-\mathbf{z}\big\|_2^2
   + \lambda\,\varphi(\mathbf{z}) \notag\\
&\quad + v\,\big\|\big(\mathbf{1}-\mathbf{u}^{(t)}\big)\odot \mathbf{z}\big\|_{1},
\end{align}
where $v\!\gg\!1$ is a penalty. Among all candidate $\mathbf{z}$, this weighted $\ell_1$ term uses $\mathbf{u}^{(t)}$ to bias the objective antenna-wise. Since $\mathbf{u}^{(t)}\!\in\![0,1]^N$, $(\mathbf{1}-\mathbf{u}^{(t)})\odot\mathbf{z}$ applies a per-antenna weight to each coefficient element-wise. For out-of-VR antennas ($u_i^{(t)}\!\approx\!0$) the weight is $v$, incurring a high cost, so that avoids attributing noise to VR energy. For in-VR antennas ($u_i^{(t)}\!\approx\!1$) the weight is near zero, yielding weak penalties. Thus, DUGC-VRNet jointly exploits antenna visibility and polar-domain sparsity through the VR-aware weighting in \eqref{deqn_ex9a} and the data-adaptive prior in \eqref{deqn_ex11a}.

\subsection{Architecture of DUGC-VRNet}

\begin{figure}[!t]
\vspace{-0.5cm}
\centering
\includegraphics[width=3.5 in]{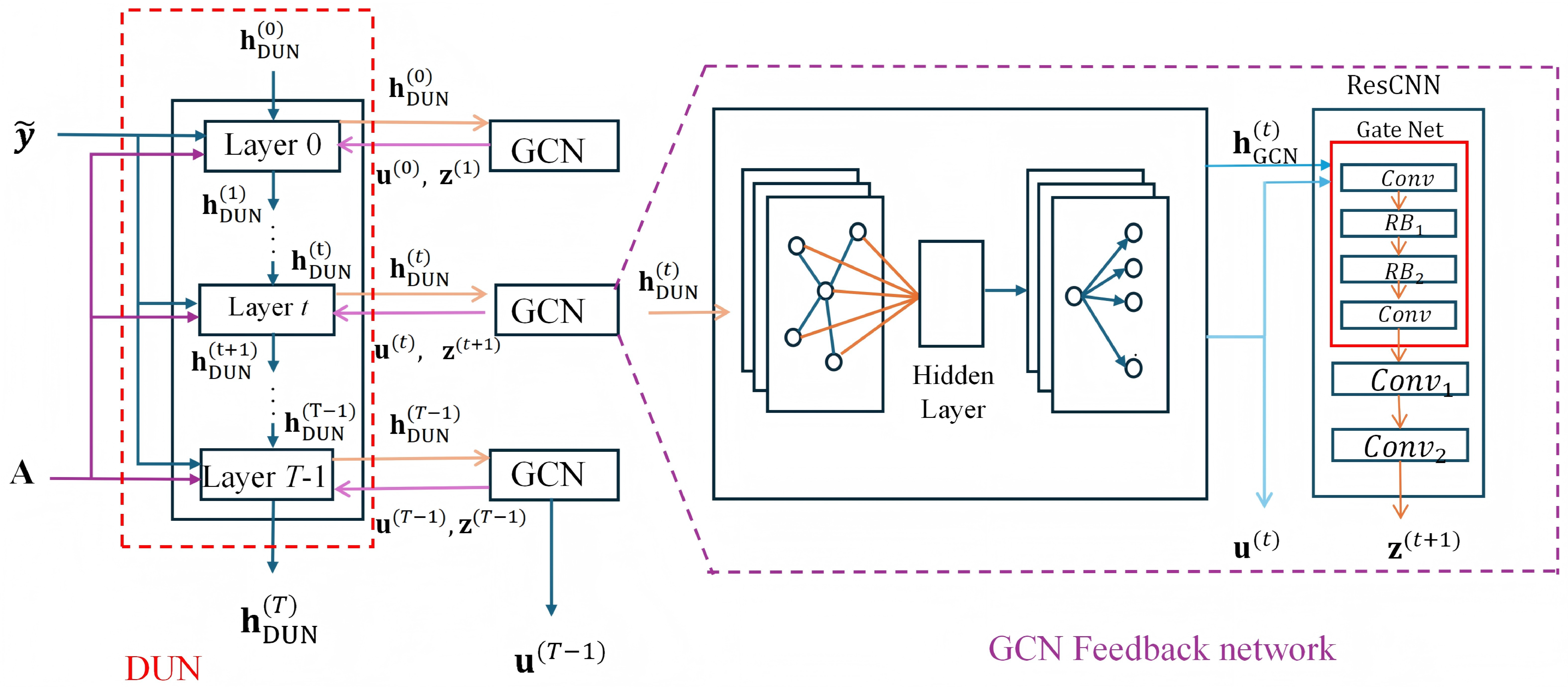}
\caption{The proposed DUGC-VRNet architecture.}
\label{fig_1}
\vspace{-0.3cm}
\end{figure}

DUGC-VRNet is designed as a $T$-layer iterative unfolding architecture to estimate the non-stationary spatial channel and recognize VR by solving \eqref{deqn_ex9a} and \eqref{deqn_ex11a}. As shown in Fig. 1, DUGC-VRNet couples a DUN with a GCN feedback network. At layer $t$, the DUN outputs an intermediate channel estimate $\mathbf{h}^{(t)}_{\mathrm{DUN}}$, and the GCN further refines it by exploiting the user-antenna graph, yielding a spatially non-stationarity-aware representation $\mathbf{h}^{(t)}_{\mathrm{GCN}}$ and a VR mask $\mathbf{u}^{(t)}$. A residual CNN (ResCNN) maps $\big(\mathbf{h}^{(t)}_{\mathrm{GCN}},\mathbf{u}^{(t)}\big)$ to $\mathbf{z}^{(t+1)}$, and $\big(\mathbf{u}^{(t)},\mathbf{z}^{(t+1)}\big)$ is fed back to the DUN to compute $\mathbf{h}^{(t+1)}_{\mathrm{DUN}}$. Each layer thus follows $\mathbf{h}^{(t)}_{\mathrm{DUN}}\!\rightarrow\!\mathbf{u}^{(t)}\!\rightarrow\!\mathbf{z}^{(t+1)}\!\rightarrow\!\mathbf{h}^{(t+1)}_{\mathrm{DUN}}$ and is trained with a joint loss function, where the GCN supplies $\mathbf{u}^{(t)}$, the ResCNN embeds spatial non\mbox{-}stationarity in $\mathbf{z}^{(t+1)}$, and the DUN uses $\big(\mathbf{u}^{(t)},\mathbf{z}^{(t+1)}\big)$ to refine the channel for higher accuracy. This $T$-layer iterative structure ensures stable convergence by combining optimization-driven descent updates with a mutually reinforcing feedback loop between the DUN and GCN, yielding a progressively decreasing loss trajectory.

\subsection{DUN for Channel Estimation}
Since $\mathbf{W}\!\big(\mathbf{u}^{(t)}\big)$ varies with $\mathbf{u}^{(t)}$ across layers, solving \eqref{deqn_ex9a} in closed form would require refactorization and matrix inversion per layer, which is computationally prohibitive. Thus, we adopt gradient descent to obtain an inexact solution \cite{ref10}. Specifically, the $t$-th layer update is derived by applying the gradient of the objective in \eqref{deqn_ex9a} as:
\begin{align}
\label{deqn_ex17a}
\mathbf{h}_{\text{DUN}}^{(t+1)}
&= \mathbf{h}_{\text{DUN}}^{(t)}
   - \gamma^{(t)} \Big[
      \mathbf{A}^{\mathrm{H}}\!\big(\mathbf{A}\mathbf{h}_{\text{DUN}}^{(t)}-\mathbf{\tilde{y}}\big)\notag\\
&\quad + \mu\,\mathbf{W}\!\big(\mathbf{u}^{(t)}\big)\big(\mathbf{h}_{\text{DUN}}^{(t)}-\mathbf{z}^{(t+1)}\big)
   \Big],
\end{align}
where $\gamma^{(t)}$ is a trainable step size, which is learned end-to-end by backpropagation within the DUN updates. 

We address \eqref{deqn_ex11a} using a data-driven ResCNN, thereby avoiding reliance on manually designed sparse dictionaries \cite{ref9}. Concretely, ResCNN applies a feature convolution with  $\operatorname{ReLU}(x)=\max (0, x)$ to enforce local smoothness, followed by a second convolution to synthesize $\mathbf{z}^{(t+1)}$:
\begin{equation}\label{deqn_ex21a}
{
\begin{aligned}
\mathbf{z}^{(t+1)}
  &= \operatorname{Conv}_2\!\Big(
      \operatorname{ReLU}\big(
        \operatorname{Conv}_1(
          \mathbf{h}_{\mathrm{GCN}}^{(t)}
          \odot
          \mathbf{m}(\mathbf{h}_{\mathrm{GCN}}^{(t)}, \mathbf{u}^{(t)})
        )
      \big)
     \Big),
\end{aligned}}
\end{equation}
where $\operatorname{Conv}_{i}$ denotes the $i$-th convolutional layer, $\mathbf{h}_{\mathrm{GCN}}^{(t)}$ is derived from $\mathbf{h}_{\mathrm{DUN}}^{(t)}$ through the GCN transformation, and a gate network reweights features:
\begin{equation}
\label{deqn_ex21a_gate_net} 
\begin{aligned}
\mathbf{m}(\mathbf{h}_{\text{GCN}}^{(t)}, \mathbf{u}^{(t)})=\operatorname{Sigmoid}\left(\mathbf{G}^{(t)}\left([\mathbf{h}_{\text{GCN}}^{(t)} \| \mathbf{u}^{(t)}]\right)\right)\in(0,1)^{N\times 1}.
\end{aligned}
\end{equation}
The Sigmoid maps the outputs of the gate network to $[0,1]$. $\mathbf{G}^{(t)}(\cdot)$ adopts a ``feature convolution $\rightarrow$ two residual blocks $\rightarrow$ feature convolution'' architecture and injects $\mathbf{u}^{(t)}$ as channel state information (CSI) by processing the concatenation $[\mathbf{h}_{\mathrm{GCN}}^{(t)} \,\|\, \mathbf{u}^{(t)}]$. During training, for the $i$-th antenna, if $u_i^{(t)} \approx 0$ (indicating it is out of the VR), the gate network drives $m_i \to 0$, so antenna $i$'s features minimally affect $\mathbf{z}^{(t+1)}$ and noise-induced bias is reduced; if $u_i^{(t)} \approx 1$ (in VR), then $m_i \to 1$, retaining the features on antenna $i$.

\subsection{GCN Feedback Network for VR Recognition}
The channel is inherently a graph structure, where the user and antennas serve as nodes, and a user–antenna edge represents a propagation path. Thus, VR is recognized using a GCN based on $\mathbf{h}_{\mathrm{DUN}}$. The GCN architecture follows an input-hidden-output paradigm. The input layer constructs a user-antenna graph with node $0$ as the user (initialized as a zero vector in $\mathbb{R}^{4 \times 1}$) and nodes $1, \ldots,N$ as BS antennas. An edge $\varepsilon_{0,n}$ is included for antenna $n$ if a potential visibility link is assumed \cite{ref11}. For antenna $n$,
\begin{equation}
\label{deqn_ex21a_x}
\begin{aligned}
\mathbf{x}_n=\big[\Re(h_{\mathrm{DUN},n}),\ \Im(h_{\mathrm{DUN},n}),\ e_n,\ p_n\big]^{\mathsf T}\in\mathbb{R}^{4 \times 1},
\end{aligned}
\end{equation}
where $h_{\mathrm{DUN},n}$ is the complex channel coefficient at antenna $n$, $\Re(\cdot)$ and $\Im(\cdot)$ denote its real and imaginary parts, $p_n=(n-1)/(N-1)$ provides absolute position information for antenna $n$ and $e_n=\sqrt{(\Re(h_{\mathrm{DUN},n}))^{2}+(\Im(h_{\mathrm{DUN},n}))^{2}}$ represents the energy at antenna $n$. Out-of-VR antennas primarily contain noise, while in-VR antennas contain both signal and noise, leading to higher total energy for the latter. We stack the node features to form the initial feature matrix
$\mathbf{X}^{(t,0)}=\big[\mathbf{x}_0^{(t)},\,\mathbf{x}_1^{(t)},\,\ldots,\,\mathbf{x}_N^{(t)}\big]^{\mathsf T}\in \mathbb{R}^{(N+1)\times F_0},
$ where $F_0=4$ is the feature dimension. Edge $(0,n)$ is presented if $e_n > \zeta{(t)}$, and absent otherwise, where $\zeta{(t)}$ is a learnable threshold and initialized by the mean energy $\zeta(0) = \frac{1}{N}\sum_{n} e_n$, and
\begin{equation}
\label{deqn_ex21a_zeta}
\begin{aligned}
\zeta(t) = \zeta(t-1) - \eta_{\zeta}\,\frac{\partial L}{\partial \zeta(t-1)}, 
\end{aligned}
\end{equation}
where $L$ denotes the loss function of DUGC-VRNet, and $\eta_{\zeta}$ is the learning rate assigned to $\zeta$ by the optimizer.

We define an adjacency matrix $\mathbf{G} \in\mathbb{R}^{(N+1)\times (N+1)}
$to store these user–antenna connections, where $g_{0,n}$ and $g_{n,0}$ are set to $1$ if $e_n > \zeta{(t)}$ and $0$ otherwise. To mitigate degree bias, we replace $\mathbf{G}$ with its normalized version $\overline{\mathbf{G}}$:
\begin{equation}
\label{deqn_ex13a}
\begin{aligned}
\overline{\mathbf{G}}=\mathbf{D}^{-\frac{1}{2}}(\mathbf{G}+\mathbf{I}) \mathbf{D}^{-\frac{1}{2}},
\end{aligned}
\end{equation}
where $\mathbf{I}$ is the identity matrix, and $\mathbf{D}$ is constructed by taking the row-wise sums of $\mathbf{G}+\mathbf{I}$ and placing them along its diagonal.

The graph convolution layers realize layer-wise propagation based on $\overline{\mathbf{G}}$, and the $l$-th layer propagation in iteration $t$ is \cite{ref7}

\begin{equation}
\label{deqn_ex14a}
\begin{aligned}
\mathbf{X}^{(t, l)}=\sigma\left(\overline{\mathbf{G}} \mathbf{X}^{(t, l-1)} \mathbf{W}_{\mathrm{GCN}}^{(l)}\right)\in \mathbb{R}^{(N+1)\times4},
\end{aligned}
\end{equation}
where $\mathbf{X}^{(t,l-1)}$ denotes the output feature matrix of the $(l - 1)$-th convolutional layer, $\mathbf{W}_{\mathrm{GCN}}^{(l)} \in \mathbb{R}^{4\times4}$ is the weight matrix of the $l$-th graph convolutional layer, and $\sigma(\cdot)$ is an activation function.

After $L$ layers, the final feature matrix $\mathbf{X}^{(t,L)}$ provides node-wise representations learned by the GCN. Since the real and imaginary parts of the channel coefficient are included in the first two components of each node feature, the refined channel estimate can be obtained through a direct data extraction process. Specifically, for each antenna node $n$, the corresponding $h_{\mathrm{GCN},n}$  is formed as a complex value using the first two entries of its feature vector in $\mathbf{X}^{(t,L)}$, which represent $\Re(h_{\mathrm{GCN},n}^{(t)})$ and $\Im(h_{\mathrm{GCN},n}^{(t)})$, respectively. After removing the user node (node $0$) and stacking the antenna-node coefficients, the resulting vector forms
$\mathbf{h}_{\mathrm{GCN}}^{(t)} \in \mathbb{C}^{N\times1},$
which is the channel information inferred by the GCN.

The intermediate VR mask vector is generated as
\begin{equation}
\label{deqn_ex15a}
\begin{aligned}
\bar{\mathbf{u}}^{(t)}=\operatorname{Sigmoid}\left(\mathbf{X}^{(t,L)} \mathbf{W}_\mathrm{GCN}^{(\mathrm{out})}\right)\in \mathbb{R}^{(N+1)\times1},
\end{aligned}
\end{equation}
where $\mathbf{W}_\mathrm{GCN}^{(\mathrm{out})} \in \mathbb{R}^{4\times1}$ is the output-layer weight vector that maps the learned node features to VR. 
Since the user node serves only as an auxiliary node for graph construction, the entry corresponding to node 0 is discarded, yielding the final VR mask vector  $\mathbf{u}^{(t)}\in \mathbb{R}^{N\times1}$.

\subsection{Multi-Task Loss Function Design}
For DUGC-VRNet, we employ a weighted multi-task loss function, defined as
\begin{equation}
\label{deqn_ex21a_L}
L \triangleq (1-\alpha)L_{\mathrm{NMSE}}(\widehat{\mathbf{h}},\mathbf{h})+\alpha L_{\mathrm{VR}}(\widehat{\mathbf{u}},\mathbf{u}),
\end{equation}
where $\alpha \in[0,1]$ is the weighting coefficient. $L_{\text {NMSE}}$ denotes the normalized mean square error (NMSE) for channel estimation, given by
\begin{equation}
L_{\mathrm{NMSE}} (\widehat{\mathbf{h}},\mathbf{h}) =
  \mathbb{E}[\|\widehat{\mathbf{h}}-\mathbf{h}\|_2^2/\|\mathbf{h}\|_2^2],
\end{equation}
where $\widehat{\mathbf{h}}$ and $\mathbf{h}$ are the predicted and true channel, respectively, and $\mathbb{E}[\cdot]$ denotes the mathematical expectation. The VR recognition accuracy term, $L_{\mathrm{VR}}$, is quantified by the expected successful detection ratio (SDR) between $\widehat{\mathbf{u}}$ and $\mathbf u$ \cite{ref4}:
\begin{equation}
L_{\mathrm{VR}} (\widehat{\mathbf{u}},\mathbf{u}) =1-\mathbb{E}\!\left[d_H(\mathbf{u},\widehat{\mathbf{u}})/l_{\boldsymbol{u}}\right].
\end{equation}
Here, $l_{\boldsymbol{u}}$ is the length of $\mathbf{u}$, $\widehat{\mathbf{u}}$ is the VR mask inferred by the GCN, and $d_H\left(\mathbf{u}, \widehat{\mathbf{u}}\right)$ denotes the Hamming distance between $\mathbf{u}$ and $\widehat{\mathbf{u}}$. In our work, $\alpha$ is set to 0.5 to balance the two tasks.

\subsection{Weight Pruning}
We reduce the model complexity via global magnitude pruning \cite{ref8}. The absolute weight $\lvert w_{ij}\rvert$ serves as the importance score, with smaller magnitudes indicating lower contribution. We gather magnitudes of the network weights from all prunable layers, sort them, and set the threshold $q$ at the $\rho$-quantile as follow,
\begin{equation}
\label{eq:rank_threshold}
q = \mathrm{Rank}\big(\{|w_{ij}|\},\, \rho\big),
\end{equation}
where $\rho$ is the pruning rate, $\mathrm{Rank}(S,\rho)$ returns the $\rho$-quantile $q$ of the multiset $S$. Let $v=\{(i,j):\,|w_{ij}|\ge q\}$. After pruning, define the binary mask $\tilde{w}_{ij}=1$ for $(i,j)\in v$ and $\tilde{w}_{ij}=0$ otherwise, thereby retaining the top $(1-\rho)\%$ of weights. The retained weights are then updated by stochastic gradient descent:
\begin{equation}
\label{eq:sgd_update}
\mathbf{W}_{\mathrm{net}}^{(s+1)}=\mathbf{W}_{\mathrm{net}}^{(s)}-\eta^{(s)}\,\nabla_{\!\mathbf{W}}L\!\big(\mathbf{W}_{\mathrm{net}}^{(s)}\big),
\end{equation}
where $\mathbf{W}_{\mathrm{net}}^{(s)}$ denotes the network weight matrix retained after pruning at iteration $s$, $\eta^{(s)}$ is the adaptive step size, and $\nabla_{\mathbf{W}} \mathcal{L}\left(\mathbf{W}_{\mathrm{net}}^{(s)}\right)$ denotes the gradient of the loss function $L$ with $\mathbf{W}_{\mathrm{net}}^{(s)}$.

\section{Simulation Results}
Computer simulations were conducted to evaluate the performance of the proposed algorithm. In the simulations, the BS is equipped with $N=256$ antennas, spaced at half-wavelength intervals, $f_{\mathrm{C}}=100$ GHz, $N_{\mathrm{RF}} = 4$ and $S = 8$. The $\theta_{\ell}$ is uniformly distributed over $\left(-\frac{\sqrt{3}}{2}, \frac{\sqrt{3}}{2}\right)$, and $r_{\ell}$ is uniformly distributed over $(4,88)$ m \cite{ref3}, Each VR mask $u_{k,s}$ is independently set to 1 or 0 with equal probability. The dataset for DUGC-VRNet comprises 16,000, 2,000, and 2,000 different channel and VR samples for training, testing, and validation, respectively. The model is trained for 100 epochs, with pruning initiated at the $50$-th epoch. The benchmark schemes for comparison are: 1) TL-OMP and LT-CEM \cite{ref3}: TL-OMP is an on-grid OMP-based scheme, while LT-CEM is an off-grid refinement method. 2) GP-SOMP performs grid-based polar-domain estimation (on-gird), and GP-SIGW refines path parameters off-grid. 3) FRM-GD \cite{ref15}: It exploits the low-rank structure of spatially non-stationary channels to estimate the channel. 4)VRDO-MP \cite{ref14}: Through Bayesian message passing, joint channel and visibility region estimation is performed. 5) MDISR-Net \cite{ref10}: This is a model-driven DUN alternating between a learnable gradient step and a CNN proximal operator for channel estimation.

Fig. 2(a) shows the NMSE performance of different schemes versus SNR with 32 pilots. The SNR is defined as $10\lg\frac{P}{\sigma_{0}^{2}}$ dB, where $\sigma_{0}^{2}$ is the average noise power normalized to 1 and $P$ is the average received signal power. VRDO-MP relies on a fixed VR ratio prior and therefore performs poorly when the visibility region is randomly generated. In contrast, the performance of other schemes improves with increasing SNR. DUGC-VRNet and its pruned variant consistently outperform all baselines. At 10 dB SNR, DUGC-VRNet achieves lower NMSE than competing methods at 20 dB. Compared with MDISR-Net, it provides about 5 dB average NMSE gain. The pruned model $(\rho=0.5)$ removes many weights with only minor degradation and still surpasses all baselines. Fig. 2(b) shows the SDR for VR recognition versus SNR. Since VR is solely determined by the on-grid estimation, with subsequent off-grid processes merely optimizing parameters like distance and angle \cite{ref3}, the SDRs of TL-OMP and LT-CEM are identical, as is the case for GP-SOMP and GP-SIGW. FRM-GD performs channel estimation well but does not model the VR, and thus cannot support VR recognition. DUGC-VRNet and its pruned variant maintain high SDR across all SNRs. Even at 0 dB, the SDR exceeds 0.9 and remains well above other schemes. For $\mathrm{SNR}\ge 5,\mathrm{dB}$, the curves nearly overlap, indicating negligible loss in VR recognition.

\begin{figure}[!t]
\centering
\begin{minipage}{0.48\linewidth}
\centering
\vspace{-0.5cm}
\subfloat[]{\includegraphics[width=1.65 in]{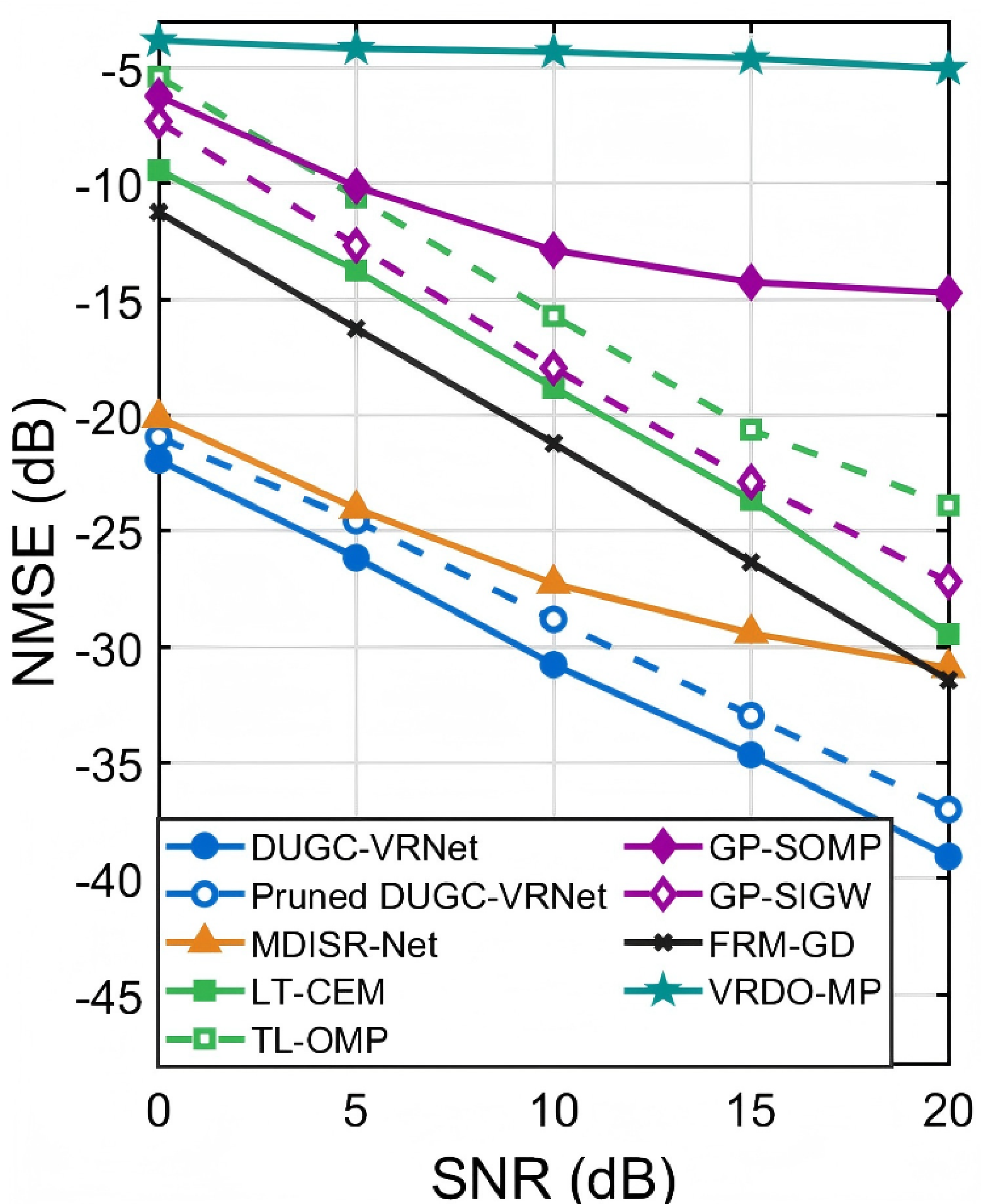}}
\end{minipage}
\hfill
\begin{minipage}{0.48\linewidth}
\centering
\vspace{-0.5cm}
\subfloat[]{\includegraphics[width=1.6 in]{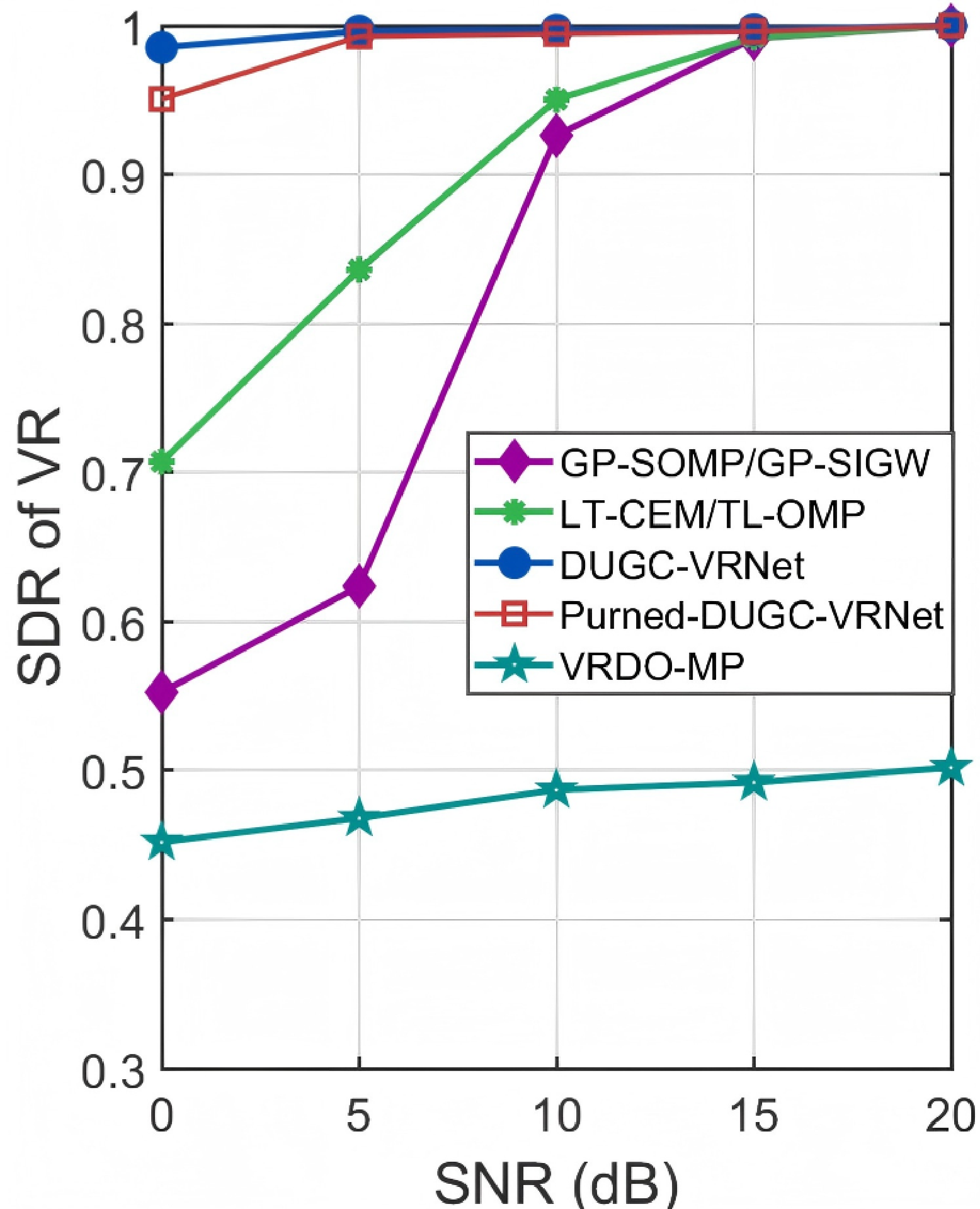}}
\end{minipage}
\caption{NMSEs and SDRs of different schemes versus SNR.}
\vspace{0.3cm}
\label{fig_2}
\end{figure}

\begin{figure}[!t]
\centering
\vspace{-0.5cm}
\begin{minipage}{0.48\linewidth}
\centering
\subfloat[]{\includegraphics[width=1.65 in]{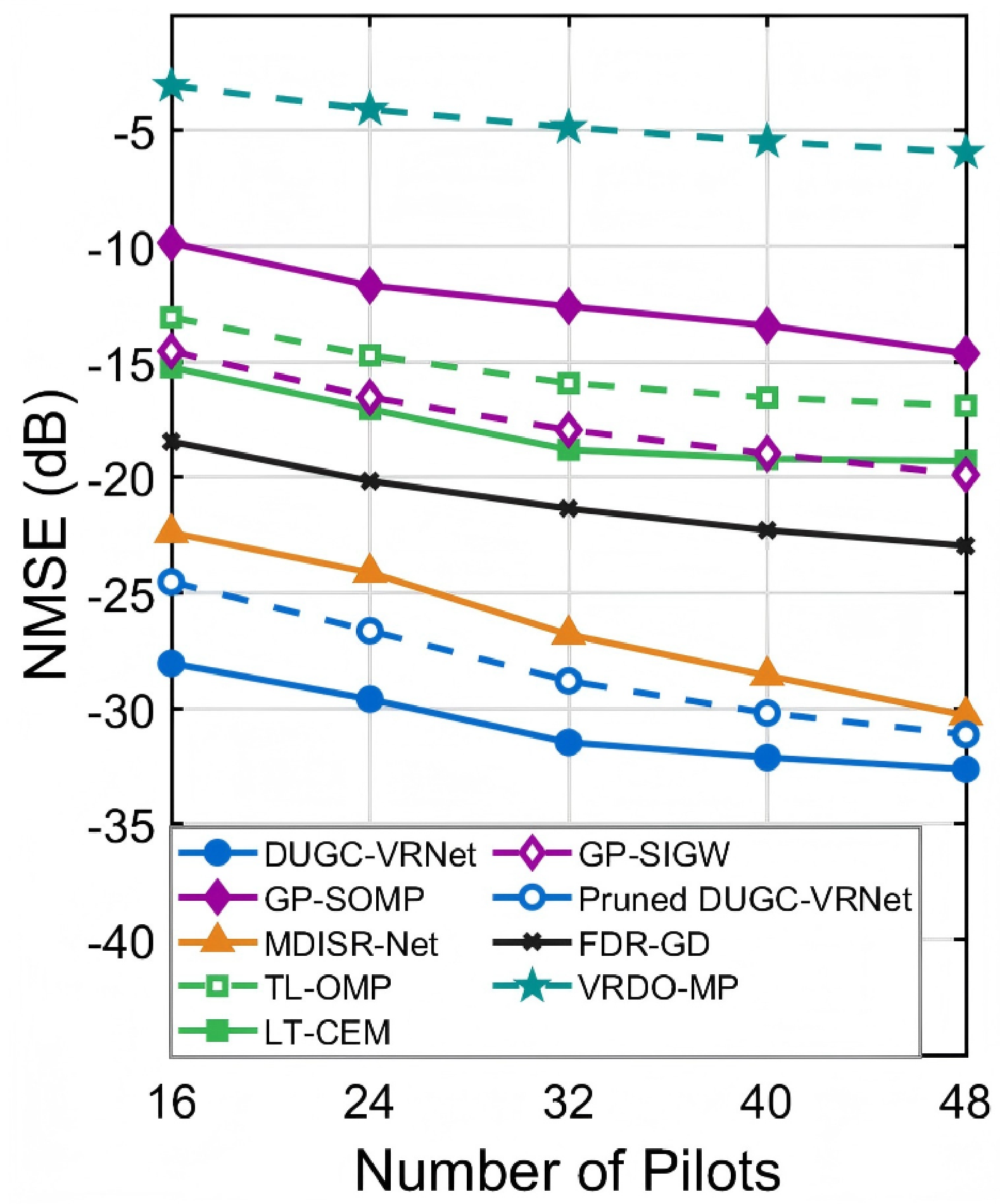}}
\end{minipage}
\hfill
\begin{minipage}{0.48\linewidth}
\centering
\subfloat[]{\includegraphics[width=1.6 in]{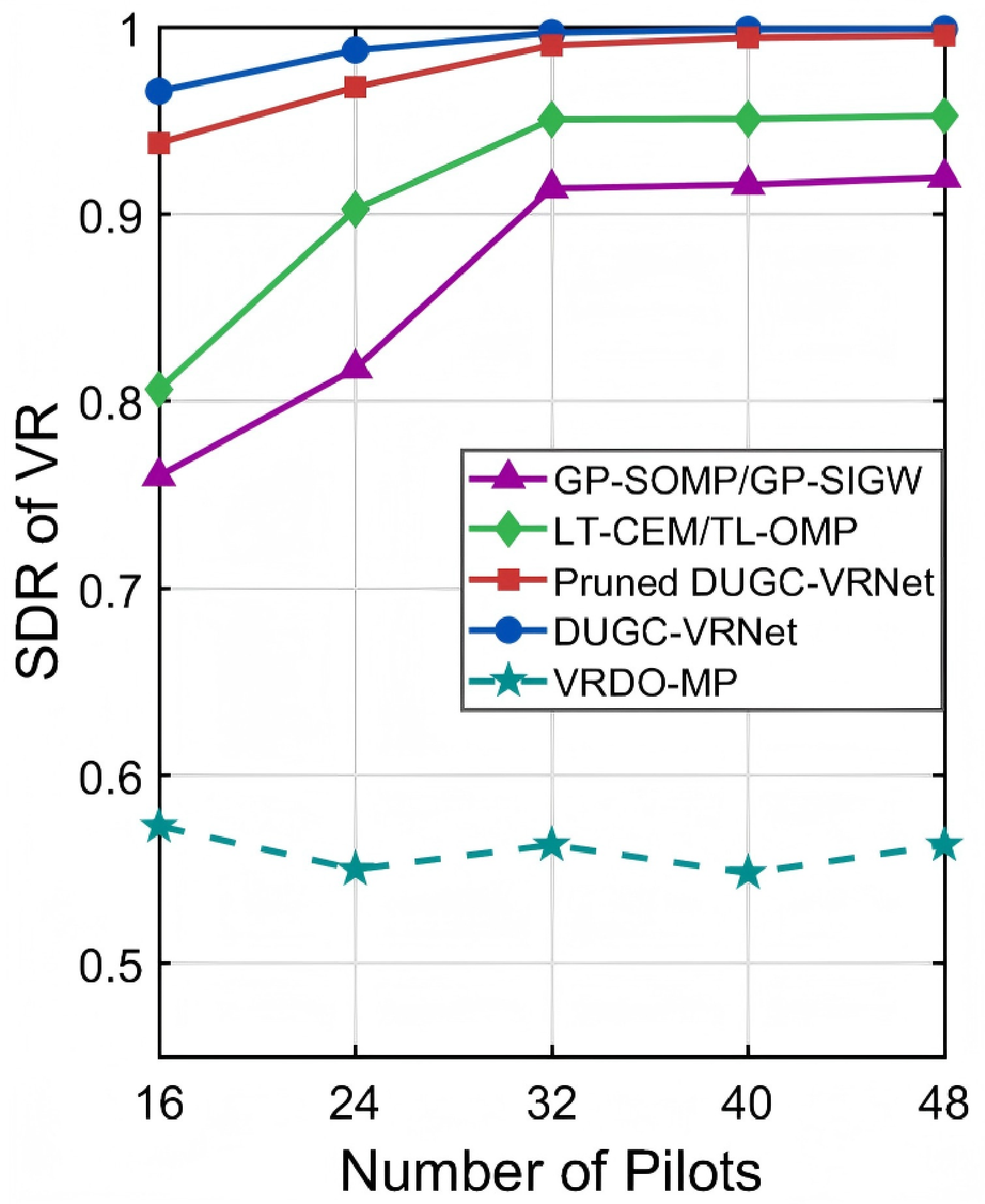}}
\end{minipage}
\caption{NMSEs and SDRs of different schemes versus SNR.}
\vspace{-0.3cm}
\label{fig_3}
\end{figure}

Fig. 3 shows the NMSE and SDR performance versus the number of pilots at SNR = 10 dB. In Fig. 3(a), DUGC-VRNet consistently outperforms all baselines and achieves comparable performance with 16 pilots to other schemes using 48 pilots. The pruned variant $(\rho=0.5)$ exhibits only minor degradation and still surpasses the competing methods, confirming effective compression. In Fig. 3(b), both DUGC-VRNet and its pruned model achieve the highest SDR across the entire pilot range, with SDR approaching one as pilots increase. The nearly overlapping curves indicate negligible impact of pruning on VR recognition.

\begin{table}
\begin{center}
\vspace{-0.5cm}
\caption{Number of parameters and performance of different schemes.}
\label{tab_params_perf}
\begin{tabular}{|c|c|c|c|c|}
\hline
Method & Params & NMSE & SDR \\ 
\hline
DUGC-VRNet & 2.59M & -31.92dB & 0.9968 \\ 
\hline
DUGC-VRNet ($\rho=0.5$) & 1.29M & -28.82dB & 0.9944 \\ 
\hline
DUGC-VRNet ($\rho=0.8$) & 516K & -23.56dB & 0.9891 \\ 
\hline
MDISR-Net & 2.41M & -27.28dB & -- \\ 
\hline
FRM-GD & -- & -21.23dB & -- \\ 
\hline
\end{tabular}
\end{center}
\vspace{-0.5cm}
\end{table}

The computational complexity of DUGC-VRNet is $\mathcal{O}\!\left(T(PN_{\mathrm{RF}}N + \mathcal{C}_{\mathrm{net}})\right)$, where $\mathcal{C}_{\mathrm{net}}$ denotes the neural network cost including the DUN, GCN and ResCNN. With global pruning, weight-dependent operations in $\mathcal{C}_{\mathrm{net}}$ are further reduced by ratio $\rho$. Table~I evaluates the impact of $\rho$ under 32 pilots and 10~dB SNR. As $\rho$ increases, the parameter count and computational load decrease while the estimation accuracy gradually degrades. At $\rho = 0.5$, DUGC-VRNet achieves MDISR-Net’s baseline accuracy with half the parameters, incurring only a 3~dB NMSE loss and negligible SDR change. A nearly five-fold parameter reduction at $\rho = 0.8$ leads to more NMSE degradation and a slight SDR drop. Nevertheless, it still outperforms FDR-GD.

\section{Conclusion}
In this letter, we propose DUGC-VRNet, which couples a DUN with a GCN to jointly perform channel estimation and VR recognition in near-field spatially non-stationary XL-MIMO. The GCN extracts VR information and feeds it back to the DUN for iterative refinement, yielding more accurate channel estimates. Simulations demonstrate superior estimation and VR accuracy, while weight pruning reduces complexity with minimal performance loss.

\end{document}